\documentclass[a4paper, 10pt, conference]{IEEEtran}
\IEEEoverridecommandlockouts                   

\usepackage{graphicx}
\usepackage{color}
\usepackage{amsmath}
\usepackage{multirow}
\usepackage{booktabs}
\usepackage{url}
\usepackage{amsmath}
\usepackage{adjustbox}
\usepackage{csquotes}
\usepackage{subfig}
\usepackage{float}
\usepackage[english]{babel}
\usepackage{comment}
\usepackage{tikz}
\usetikzlibrary{positioning}

\newcommand\copyrighttext{%
  \footnotesize \textcopyright 2023 IEEE. Personal use of this material is permitted. Permission from IEEE must be obtained for all other uses, in any current or future media, including reprinting/republishing this material for advertising or promotional purposes, creating new collective works, for resale or redistribution to servers or lists, or reuse of any copyrighted component of this work in other works. M. Vergari, T. Kojic, N. S. Bertges, F. Vona, S. Möller and J. -N. Voigt-Antons, "Exploring users' sense of safety in public using an Augmented Reality application," 2023 15th International Conference on Quality of Multimedia Experience (QoMEX), Ghent, Belgium, 2023, pp. 193-196, doi: 10.1109/QoMEX58391.2023.10178675. https://ieeexplore.ieee.org/document/10178675}
\newcommand\copyrightnotice{%
\begin{tikzpicture}[remember picture,overlay]
\node[anchor=south,yshift=10pt] at (current page.south) {\fbox{\parbox{\dimexpr\textwidth-\fboxsep-\fboxrule\relax}{\copyrighttext}}};
\end{tikzpicture}%
}

\begin{document}

\title{Exploring users' sense of safety in public using an Augmented Reality application}


\author{
 \IEEEauthorblockN{Maurizio Vergari$^1$, Tanja Koji\'c$^1$, Nicole Stefanie Bertges $^1$, Francesco Vona $^3$,\\ Sebastian M\"oller$^{1,2}$, Jan-Niklas Voigt-Antons$^3$}
 \IEEEauthorblockA{$^1$Quality and Usability Lab, TU Berlin, Germany \\  $^2$German Research Center for Artificial Intelligence (DFKI), Berlin, Germany\\ $^3$Immersive Reality Lab, Hamm-Lippstadt University of Applied Sciences}
 
}


\maketitle
\copyrightnotice

\begin{abstract}
Nowadays, Augmented Reality (AR) is available on almost all smartphones creating some exciting interaction opportunities but also challenges. For example, already after the famous AR app Pokémon GO was released in July 2016, numerous accidents related to the use of the app were reported by users. At the same time, the spread of AR can be noticed in the tourism industry, enabling tourists to explore their surroundings in new ways but also exposing them to safety issues. This preliminary study explores users' sense of safety when manipulating the amount and UI elements visualization parameters of Point of Interest (POI) markers in a developed AR application. The results show that the amount of POI markers that are displayed is significant for participants' sense of safety. The influence of manipulating UI elements in terms of \enquote{transparency}, \enquote{color}, and \enquote{size} cannot be proven. Nevertheless, most tested people stated that manipulating transparency and size somehow influences their sense of safety, so a closer look at them should be taken in future studies. 
\end{abstract}

\begin{keywords}
    Augmented Reality, Safety, User Experience Design, Visualization
    \vspace{1em}
\end{keywords}


\section{INTRODUCTION \& RELATED WORK}
In the last years, Augmented Reality (AR) experienced increasing popularity. As AR is nowadays available for smartphone usage, the number of applications exploiting AR increased, and it has become a technology for the mass market. At the same time, the tourism market is increasingly being affected by the advent of new technologies \cite{buhalis2019}. It is not surprising that nowadays, tourists explore unknown environments using their smartphones instead of a guidebook. 

Unfortunately, it needs to be considered that there are many accidents related to smartphone usage on the street \cite{nasar2013}. In a study by the Pew Research Center in 2014, more than half of the adult smartphone users reported that they had already experienced collisions with other pedestrians due to smartphone distraction \cite{smith2014}. The same problem can be found in AR applications for tourism, where people have to look at their smartphone screens to interact with a Point of Interest (POI). Thus AR applications can be very engaging, but at the same time, this poses a risk regarding smartphone usage \cite{Rovira2020}\cite{Olsson2011}. Navigation is also another important area where AR is applied. AR navigation aids have been designed and used mostly for pedestrian way-finding. Such AR navigation systems do not provide an overview of the entire route. The way to the destination is shown from the own point of view using symbols and other visual information. For Dong et al., this type of representation \enquote{expresses geographical information and relationships more specifically} \cite{Dong2021} and reduces cognitive effort, allowing people to pay more attention to others around them compared to using a 2D digital/virtual map. Unnecessary information does not need to be displayed if the corresponding object is undetected or irrelevant. However, this requires precise placement of the augmented information in the real world. 
Aside from the importance of accurate information placement, little research has been done on the User Interface of AR tourism applications, especially regarding hand-held devices.
The main problem is the visualization and positioning of the virtual objects \cite{Yovcheva2012}. Depending on the number, size, and visualization options, large display parts can also be cluttered with virtual objects in an AR app.
Without filtering information, the relatively small display of a smartphone is quickly overloaded, resulting in an obstruction of view of sights or even traffic, pedestrians, and other objects. Researchers mentioned limiting the displayed content according to the distance between the user and the POI as a possible solution. Others do not consider general filtering the solution, as the users may want to discover the unfamiliar surroundings without pre-filtering \cite{ajanki2011, julier2000, tang2003, Yovcheva2014}.
Olsson et al. \cite{Olsson2011} name the importance of showing information sensitive to the context. A simple layout with a balanced integration of virtual cues and reality is also required \cite{bloksa2017}. For example, to make the best decision regarding the color scheme of the annotations, some research suggests using billboard style. Here, the background color changes according to the real environment \cite{jankowski2010, gabbard2008}. Moreover, Bell et al. \cite{bell2001} suggested placing information automatically so that these do not occlude each other. 

All the areas above could affect users' sense of safety. Safety is the state of being safe and protected from danger or harm. In this paper, the concept of sense of \enquote{safety} is defined somewhat broader, meaning to have a feeling of enough cognitive workload left to be aware of the surroundings and to be able to direct sufficient attention to move away from potential obstacles and dangers. These hazards can be static as well as dynamic with moving objects or people. In addition, not only the own safety but also other peoples' and objects' should be considered safety-relevant. As Rovira et al. pointed out, AR applications can be very engaging, and this is a desired effect in other digital media but a risk when it comes to AR on hand-held devices such as smartphones \cite{Rovira2020, Olsson2011}. Additionally, they measured in their work that participants spent, on average, 86\% of time looking at their smartphone screen while walking \cite{Rovira2020}. Safety in terms of distraction and awareness of things and people around has been barely researched, especially for pedestrian safety. However, when designing mobile AR applications, safety should be taken into account. 

\subsection{Objectives} 

The general objective of this study is to investigate how different variables influence the users' sense of safety when using AR tourism applications. As shown by the presented related work, the usage of smartphones on the streets may pose risks for oneself and others. The expanding market of mobile AR requires research to increase users' safety. Specifically, variations in size, transparency, color, and amount of POI marker options have been evaluated. The following two research questions arise:

\begin{itemize}
    \item How does the amount of shown information points influence the users' sense of safety?
    \item How do visual parameters (transparency, color, size) influence users' sense of safety?
\end{itemize}

\section{METHODS}
\subsection{Design and Test setup}

The developed application aims to measure values for visual parameters considered important for the UI of an AR experience. Therefore, the application allows to adjustment of values for transparency, color, and size and shows different amounts of POI markers. All these elements are considered relevant to users' sense of safety, as they can alter users' perception of the real environment. The application was developed using Unity and AR Foundation. The test device was an Android device - the Samsung Galaxy S7. The smartphone's display is 5.1 inches (130 mm). The graphical user interface mainly consists of the sliders area (see Fig. \ref{fig:UI}). This allows adjusting values for transparency, color, and size. Besides the sliders, there are only two unobtrusive buttons in the upper corners that allow hiding the GUI and saving the values. Different POI markers were placed in fixed positions and evenly on a length of about 50 meters in each condition. Their positioning is based on a real street map downloaded from \enquote{OpenStreetMap.org}. In total, nine conditions are given by the combination of the two independent variables \enquote{UI-Element} and \enquote{Amount} of the POI markers. The independent variables had three levels each: 

\begin{figure}
\centering
    \subfloat{
    \includegraphics[scale=0.25]{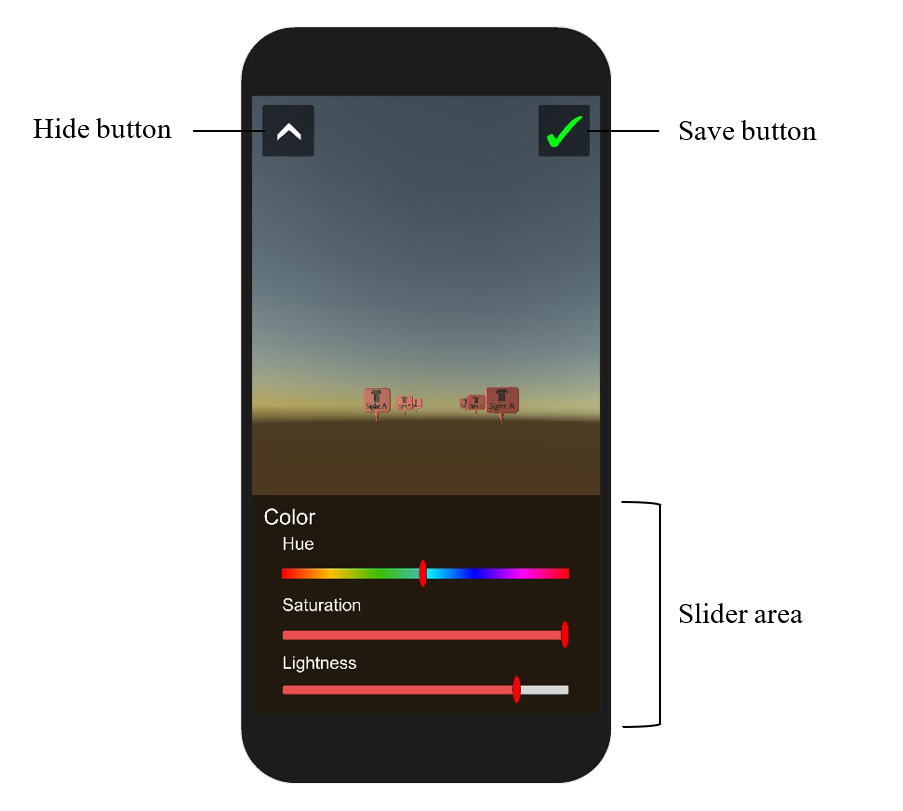}
    }
    \subfloat{
    \includegraphics[scale=0.07]{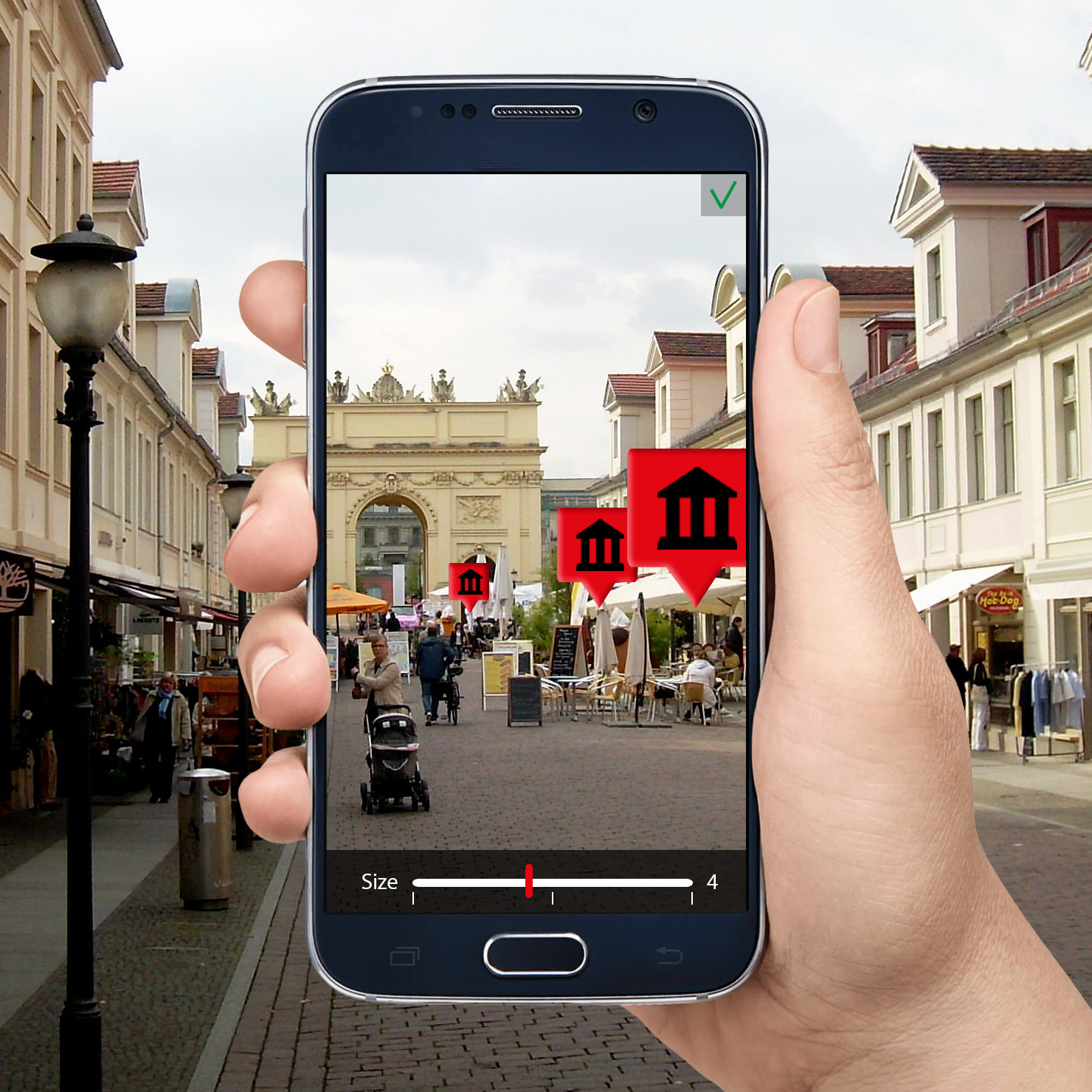}
    }
    
    \caption{Examples of the Application GUI}
    \label{fig:UI}
\end{figure}

\textbf{UI-Element:} \textit{Transparency slider}: the transparency of the POI marker can be set using the slider (0-1); \textit{Color sliders:} the three sliders can be used to manipulate color in terms of Hue, Saturation, and Lightness; \textit{Size sliders}: the two sliders can be used to manipulate height and width of the POI marker.

\textbf{Amount :} \textit{Few Markers:} three markers placed per roadside occupy about 55\% of the screen width; \textit{Medium amount of Markers:} five markers placed per roadside occupying 60\% of screen width; \textit{Many Markers:} seven markers placed per road sides occupying 65\% of screen width.

\subsection{Procedure}

A group of 25 people took part in the study. They were 15 men and 10 women. The average age was 28.84 (SD=9.31, min=20, max=59). The user study for this research was done in a main public street of (removed - double-blind), where the participants were invited, each at a different time slot. The moderator gave them an introduction to the topic and tasks. The main part of the study was the gameplay of 9 conditions, combining 3 UI-Elements (transparency, color, and size) with 3 different amounts of markers (few, medium, many). The participants were asked to provide some general information starting with demographics. Also, as part of the pre-questionnaire, the participant's tendency to generally engage with technology was recorded using the Affinity for Technology Interaction (ATI) Scale questionnaire \cite{franke2019personal}. The average measured ATI score was 4.03 (SD=1.001). To investigate the general safety related to smartphone behavior of the participant, a modified version of the Problematic Mobile Phone Use Questionnaire (PMPU-Q) was used, including only the danger dimension  \cite{billieux2008,kuss2018}(see Table \ref{Problematic Smartphone Use Questionnaire - Updated Version}). The individual questions were adapted to address pedestrians rather than car drivers' usage. The measurement of the modified PMPU-Q resulted in a mean value of 2.28 (SD=0.566). This value is in the middle range of the scale of the PMPU-Q. High scores suggest problematic, dangerous use. 

   \vspace{-1em}
   \begin{table}[H]
    \centering
        \caption{Problematic Smartphone Use Questionnaire - Modified version}
        \label{Problematic Smartphone Use Questionnaire - Updated Version}
        \resizebox{\columnwidth}{!}{
            \begin{tabular}{ll}
            \hline
            1. I use my mobile phone while \textit{walking}.*\\  \hline
            2. I try to avoid using my mobile phone when \textit{walking on the street}.\\  \hline
            3. I use my mobile phone in situations that would qualify as dangerous.* \\  \hline
            4. While \textit{walking}, I find myself in dangerous situations because of my \\ mobile phone use.*\\  \hline
            5. I use my mobile phone while \textit{walking}, even in situations that require \\ a lot of concentration.*\\  \hline
        \end{tabular} 
        }
        \vspace{-2em}
    \end{table}

After each task, the following questionnaires were given to participants to measure the effect of each condition:

Smartphone Distraction Scale (SDS): The scale is used to evaluate how much people are distracted by their smartphones \cite{throuvala2021}. In this study, only the \enquote{Attention Impulsiveness} factor was of interest and therefore adapted (see Table \ref{Smartphone Distraction Scale - Modified Version}). Higher values indicate more distraction.

Smombie Scale: This questionnaire was designed to measure pedestrians' smartphone use and help prevent dangerous behavior and to deal with risks \cite{park2021}. In this study, only the \enquote{Perceived risk} factor was used and modified (see Table \ref{Smombie Scale - Modified Version}). Higher values indicate a higher perceived risk.  

\begin{table}[H]
        \centering
        \caption{Smombie Scale - Modified version}
        \label{Smombie Scale - Modified Version}
        \resizebox{\columnwidth}{!}{
        \begin{tabular}{ll}
    
            \hline
            1. I think when using the app while walking outside, it could \\ cause a traffic crash. \\  \hline
            2. I think when using the app while walking outside, it would take me \\ longer to notice a bicycle or car. \\  \hline
            3. I think when using the app while walking outside, I could bump \\ into another person.\\ \hline
            4. I think when using the app while walking outside, I could miss \\ a crosswalk signal.\\  \hline
            5. I think when using the app while walking outside, I might miss \\ an obstacle on my path. \\  \hline
            6. I think when using the app while walking outside, I wouldn't notice \\ if someone is trying to get my attention. \\  \hline
        \end{tabular} 
        }
        \vspace{-1em}
   \end{table}
   
Short User Experience Questionnaire (UEQ-S): It is a questionnaire to measure users' subjective impressions of products' user experience \cite{schrepp2017design}. 

Self-Assessment Manikin (SAM): It is an emotion assessment tool that uses graphic scales depicting cartoon characters expressing three emotional elements: valence, arousal, and dominance \cite{bradley1994measuring}.

After testing all conditions, the participant completed the System Usability Scale (SUS) \cite{brooke1996sus}. The tests took around 60 minutes on average. The conditions were run using Latin square randomization to avoid sequence effects \cite{LatinSquare}.

\section{RESULTS}
A repeated measure Analysis of Variance (ANOVA) was run to detect statistically significant differences. Table \ref{tab:results} provides an overview of the significant effects found.

\begin{table}[htbp]
  \centering
  \caption{Effects of different amount of Markers (Amount) on modified SDS and Smombie Scale}
  \resizebox{\columnwidth}{!}{
    \begin{tabular}{lllcllrl}
    \toprule
\multicolumn{1}{l}{Effect} & \multicolumn{1}{l}{Parameter} & \multicolumn{1}{c}{$df_{\textnormal n}$} & \multicolumn{1}{l}{$df_{\textnormal d}$} & \multicolumn{1}{c}{$F$} & \multicolumn{1}{c}{$p$} & \multicolumn{1}{c}{$\eta_{\textnormal G}^2$}
    \\
    \midrule
    Amount &  Safety\_SDS & $2$     & $48$    & $ 18.674$ & $ <0.001$ & $0.438$ \\	
    Amount &  Safety\_Smombie & $2$     & $48$    & $ 11.663$ & $ <0.001$ & $0.327$ \\
    \bottomrule
    \end{tabular}%
    }
  \label{tab:results}%

\end{table}%

\begin{figure}
    \centering
    \subfloat{
    \hspace{-1em}
    \includegraphics[scale=0.24]{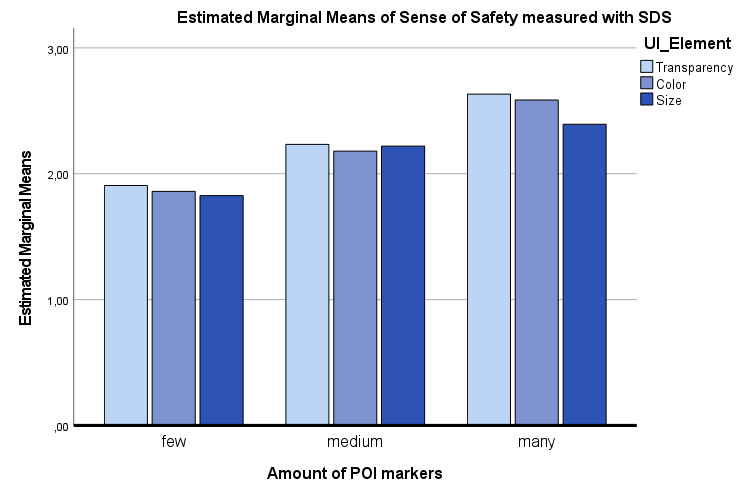}
    }
     \subfloat{
     \hspace{-1.5em}
    \includegraphics[scale=0.24]{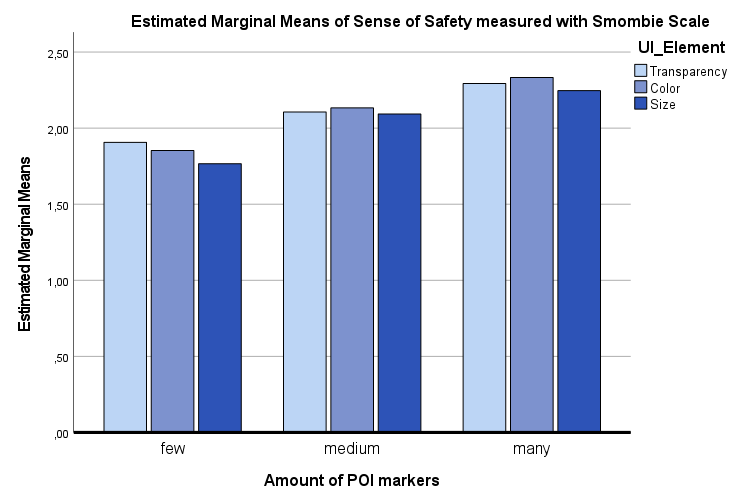}
    }
    
    \caption{Estimated Marginal Means for modified SDS (on left) and Smombie Scale (on right) for amount of markers and UI Element}
    \label{fig:results}
    \vspace{-2em}
\end{figure}

Overall, the results show a significant difference in users' sense of safety regarding the amount of POI markers. Significant results were found for the modified SDS questionnaire (see Fig. \ref{fig:results}). It turns out that a higher amount of POI marker leads to a higher mean value (Few: M=1.864, SE=0.141; Medium: M=2.211, SE=0.136; Many: M=2.538, SE=0.135). The pairwise comparison with Bonferroni adjustment resulted in significance for each pair of levels. A similar trend is noticeable (see Fig. \ref{fig:results}) from the modified Smombie Scale questionnaire (Few: M=1.842, SE=0.162; Medium: M=2.111, SE=0.17; Many: M=2.291, SE=0.152). The pairwise comparison with the Bonferroni adjustment shows no significant difference in users' sense of safety between the amount levels medium (M=2.111, SE=0.17) and many (M=2.291, SE=0.152).
Based on the questionnaires, a difference in manipulating the UI elements in separated conditions on Sense of Safety was not noticeable.


From the analysis of the SUS \cite{brooke1996sus} results, the developed AR experience was rated with a score of 86 (SD = 6.20). The score indicates that the application usability is above average ($>$68), indicating a good usability level.


\section{DISCUSSION \& CONCLUSION}
This research aimed to explore how different variables influence the users' sense of safety when using AR tourism applications.
After investigating the independent variables \enquote{UI-Element} and \enquote{Amount} of the POI markers, the results determined the influence of the amount of POI markers on the variable \enquote{Sense of Safety}. The SDS score showed an influence on the \enquote{Sense of Safety} variable for different amounts of POI markers. While the Smombie Scale was only significant for the pairwise comparison of the few-medium and few-many levels. Results indicate no significant difference between level medium and many, but the average value for the medium amount of markers is still slightly lower than the many markers. Therefore, a general influence of the amount of POI markers can be assumed. Fewer markers make users feel safer. The decreasing value of the variable \enquote{Sense of Safety} for an increasing amount of POI markers suggests that a high amount of markers makes the display look cluttered and prevents an unobstructed view. This explanation is supported by some literature that shows cluttering displays and overwhelming AR cues as risky \enquote{distractions to relevant cues of the physical environment} \cite{VanKrevelen_Poelman2010, Olsson2011, tang2003}. 

When interacting with the UI elements, the users were asked to adjust those elements to maximize their sense of safety. The levels of \enquote{UI-Element} were 1) transparency, 2) color, and 3) size. A significant influence could not be measured using the questionnaires. This is probably due to changing lighting conditions because of weather and shadowed and sunny areas, which differ while walking in the pedestrian zone. In addition, the background (e.g., buildings, pedestrians) changed while experimenting and could have influenced participants' choices \cite{gabbard2008}. In addition, 76\% of the participants reported that they think there is a difference in their sense of safety when using different values for transparency. 88\% reported such an influence for the size.

This study was a limited to a single public setting; therefore a possible extension would be to conduct field experiments in various public settings with a wider variety of spectators and different environmental elements. Also, varying noise levels for a more authentic representation and testing it with a head-mounted AR device or varying the app's kind of content could bring interesting contributions to the topic. Finally, our study suggested taking a closer look at the effect of UI element transparency and size should be considered in future studies. 

\bibliographystyle{IEEEtran}
\bibliography{main.bib} 

\end{document}